\documentclass[9pt, twocolumn]{extarticle}
\usepackage{authblk}
\usepackage{lmodern}
\usepackage[scaled]{helvet}
\usepackage[T1]{fontenc}
\usepackage[twoside,%
				letterpaper,includeheadfoot,%
				layoutsize={8.125in,10.875in},%
                layouthoffset=0.1875in,%
                layoutvoffset=0.0625in,%
                left=38.5pt,%
                right=43pt,%
                top=43pt,%
                bottom=32pt,%
                headheight=0pt,%
                headsep=10pt,%
                footskip=25pt,
                marginparwidth=38pt]{geometry}
\usepackage{graphicx,amsmath,eqnarray,amssymb}
\usepackage[outdir=./]{epstopdf}
\usepackage{booktabs}
\usepackage{etoolbox}
\usepackage[flushmargin,ragged,symbol*]{footmisc}
\AtBeginEnvironment{tabular}{
\sffamily\fontsize{7.5}{10}\selectfont
}

\providecommand{\keywords}[1]
{
  \small	
  \textbf{\textit{Keywords---}} #1
}

\def\s{\mathbf{s}}
\def\r{\mathbf{r}}
\def\hs{\hat{\mathbf{s}}}
\def\hr{\hat{\mathbf{r}}}

\def\un{^{(n)}}
\def\bsigma{\boldsymbol{\sigma}}

\usepackage{xr}
\usepackage[labelfont={bf,sf},%
                labelsep=period,%
                figurename=Fig.]{caption}
\DeclareCaptionFormat{customformat}{\normalfont\sffamily\fontsize{7}{9}\selectfont#1#2#3}
\title{Evaluating Evasion Strategies in Zebrafish Larvae}
\author[a]{Yusheng Jiao}
\author[a,b]{Brendan Colvert} 
\author[a,c]{Yi Man}
\author[d]{Matthew J. McHenry}
\author[a,*]{Eva Kanso\thanks{To whom correspondence should be addressed. E-mail: kanso@usc.edu}}

\affil[a]{Aerospace and Mechanical Engineering, University of Southern California, 854 Downey way, Los Angeles, California 90089, USA}
\affil[b]{Department of Bioengineering, University of California, San Diego, 9500 Gilman Dr, La Jolla, CA 92093, USA}
\affil[c]{Department of Mechanics and Engineering Science at College of Engineering and LTCS,  Peking  University,  Beijing  100871,  P.  R.  China.} 
\affil[d]{Department of Ecology and Evolutionary Biology, University of California, Irvine, 321 Steinhaus Hall, Irvine, CA 92697, USA}

\begin{document}

\maketitle
\begin{abstract}
An effective evasion strategy allows prey to survive encounters with predators. 
Prey are generally thought to escape in a direction that is either random or serves to maximize the minimum distance from the predator.
Here we introduce a comprehensive approach to determine the most likely evasion strategy among multiple hypotheses and the role of biomechanical constraints on the escape response of prey fish. 
Through a consideration of six strategies with sensorimotor noise and previous kinematic measurements, our analysis shows that zebrafish larvae generally escape in a direction orthogonal to the predator's heading. By sensing only the predator's heading, this orthogonal strategy maximizes the distance from fast-moving predators, and,
when operating within the biomechanical constraints of the escape response, it provides the best predictions of prey behavior among all alternatives. This work demonstrates a framework for resolving the strategic basis of evastion in predator-prey interactions, which could be applied to a broad diversity of animals.
\end{abstract}\hspace{10pt}

\noindent{\footnotesize Author contributions: EK secured funds and designed and supervised research. YJ, BC, YM, and EK performed research. YJ, BC, YM, MJM, and EK analyzed results. YJ, YM, and EK wrote the paper and YJ, BC, YM, MJM, and EK revised and edited it.}

\noindent{\footnotesize Author declaration: The authors declare no conflict of interest.}
\section*{Introduction}
The abilities to sense and evade predators are central to the survival of a diversity of prey species.
The timing, speed, and direction of a prey's escape reflect the animal's evasion strategy, which is formulated by its neurophysiology and biomechanics~\cite{Schall1980}.
Despite the fundamental importance of predator encounters, resolving a prey's strategy is experimentally challenging due to the variability inherent to animal behavior.
Predators vary in their approach toward prey and the ability of the prey to respond is filtered through the environment and the animal's physiology, which may additionally introduce noise in sensing, integration, and motor response.
The aims of the present study are to develop an analytical approach that is capable of resolving prey strategy from kinematic measurements and to use that approach to test classic theory on strategy in fish predator-prey interactions.

The interactions between an individual predator fish and prey fish offer a classic system for the study of evasion strategy.
Fish evade predators with a stereotypical `C-start' response, characterized by the fish body bending into a preparatory ‘C’ shape, followed by a rapid acceleration as the body unfolds with largely planar motion~\cite{Weihs1973}. 
Fish escape behavior inspired an application of differential game theory to determine the optimal strategy of prey~\cite{Weihs1984}.
The \textit{distance-optimal strategy} is the solution to the `homicidal chauffeur' game where prey move in the direction that maximizes the closest distance achieved by a predator that maintains a constant velocity~\cite{Isaacs1999}. 
The \textit{distance-optimal strategy} has been invoked to explain the 
escape responses in animals as divergent as cockroaches \cite{Domenici2008}, crickets \cite{Casas2014}, shrimp \cite{Arnott1999}, frogs \cite{Brown1995}, salamanders \cite{Azizi2002}, crabs \cite{Woodbury1986}, and a variety of fish species \cite{Trujillo2022,Domenici2002}.
This strategy is generally considered the primary alternative to escaping in a random direction, known as the \textit{protean strategy}, which offers the tactical benefit of confusing the predator~\cite{Humphries1970, Driver1988, Moore2017, Domenici2011b}.
The present study considers whether previously-measured escape kinematics in zebrafish larvae~\cite{Nair2017,Stewart2014} are consistent with distance-optimal, pure-protean, or alternative strategies. 

The zebrafish (\textit{Danio rerio}) larva is a compelling system for investigating  evasion because it has served as a model for the neurophysiology and biomechanics of the C-start.
Its small size, lack of pigmentation, and amenability to genetic manipulation have facilitated applications of functional imaging and optogenetics in zebrafish to observe and manipulate the sensory and motor circuits responsible for visually-mediated escapes~\cite{Dunn2016, Lacoste2015, Higashijima2003}.
A combination of high-speed kinematics, flow visualization, and computational fluid dynamics have revealed a comprehensive accounting of the fluid forces that propel the escape response of zebrafish larvae~\cite{Stewart2013, Voesenek2019, voesenek2018, Muller2004, Gazzola2012}.
We incorporate these findings into a consideration of the biomechanical constraints on the escape strategy.

\begin{figure*}[!t]
\centering\includegraphics[scale = 1]{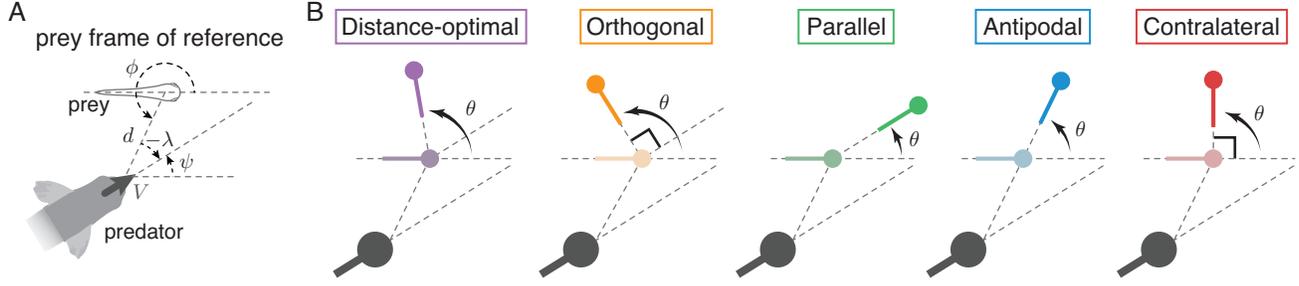}
\caption{Evasion strategies. (A) Schematic shows the predator position $(d,\phi)$ and heading $\psi$ in the prey's frame of reference. 
(B) The change $\theta$ in prey heading direction at evasion as predicted by five evasion strategies: 
\textit{distance-optimal}, prey makes a turn that maximizes the shortest distance from predator; \textit{orthogonal}, prey turns to the direction orthogonal to the predator heading in order to flee the path of the predator; \textit{parallel}, prey turns to align with the predator heading direction; 
\textit{antipodal}, prey turns in the opposite direction of the predator angular position; and 
\textit{contralateral}, prey turns left or right by 90$^\circ$ depending on the predator angular position. Strategies are distinguished by color.}
\label{fig:strategies}
\end{figure*}

We adopt a multi-pronged approach for testing the evasion strategy in larval zebrafish.
Using a strong-inference technique~\cite{Platt1964}, we mathematically define models for six strategies, both with and without sensorimotor noise (Fig.~\ref{fig:strategies}). 
We then proceed to evaluate these model predictions against previous measurements of escape kinematics~\cite{Nair2017} to determine the strategy that most-likely describes those observations.
Finally, a consideration of the escape hydrodynamics and fluid-structure interactions allows us to evaluate the constraints on these strategies.
These measures combine to offer a general framework for evaluating evasion strategies in predator-prey encounters.

\section*{Results}

Our description of the major results is organized around four main themes: (1) experimental data of the evasion kinematics of larval zebrafish and their descriptive statistics, (2) mathematical definitions of the evasion strategies, (3) formulation of the analytical approach and testing of evasion strategies, with and without sensorimotor noise, and (4) evaluation of the effects of biomechanical constraints on evasion.

\subsection*{Experimental measurements of escape kinematics}

We analyzed anew a large experimental dataset for the kinematics of escape responses in zebrafish larvae that were previously published~\cite{Nair2017,Stewart2014}. 
Larvae were exposed to a robotic predator, consisting of a sacrificed adult zebrafish (of fixed size) controlled with a motor to move through an aquarium of otherwise still water.
As detailed previously~\cite{Nair2017,Stewart2014}, larvae were largely motionless prior to the escape response that was stimulated by the presentation of the predator. 
The recorded responses of larvae were compiled from numerous experiments, each of which elicited a modest number of responses.
Larvae were excluded from the analysis if they responded within a few body lengths, or a few seconds after, another responding larva.
The 3D kinematics of larvae were compiled in the predator's frame-of-reference to yield a cloud of responses anterior to the robotic predator.

The speed of the robotic predator was set to a constant equal to $2$, $11$, or $20$  cm$\cdot$s$^{-1}$ to reflect the speed range of a typical foraging predator~\cite{Stewart2013}. This ensured a repeatable stimulus that elicited a fast C-start response from the larvae~\cite{Stewart2014,Nair2017}. 
High-speed kinematics recorded a total of $699$ evasion instances: $N_\textrm{slow} = 251$ for the slow-moving predator, 
$N_\textrm{mid} = 233$ for the mid-speed predator, 
and $N_\textrm{fast} = 215$ for the fast-moving predator (Fig.~\ref{fig:expData}).

\subsubsection*{Experimental analysis} 

From the previous kinematic measurements, we presently calculated the predator distance $d$, angular position $\phi\in[0,2\pi)$, and heading $\psi\in[-\pi,\pi)$ in the prey's frame of reference at the onset of the C-start escape response, and we calculated the change in the prey's orientation $\theta\in[-\pi,\pi)$ as it completed the C-start escape response (Fig.~\ref{fig:strategies}A and SI, Fig. S1B).
In our analysis, $\theta$ captures the rotation of the entire fish body, that is, the change in prey heading, which is not the same as the change in the body angular position employed previously~\cite{Nair2017,Stewart2014}.
We clearly distinguish between the prey's sensing of the predator angular position $\phi$ and heading $\psi$, which are often confused in empirical studies of evasion~\cite{Domenici1993,Eaton1991,Domenici2002}. 
In addition to the predator's actual heading direction $\psi$, we considered that the prey perceives $\lambda$, the deviation of the predator's heading from the angular position $\phi$, given by $\lambda = \psi - (\phi +\pi)$, $\lambda\in[-\pi,\pi)$ (see  Fig.~\ref{fig:strategies}A and SI, Fig. S1B, C). 
The predatory stimulus is said to be
sinistral if $\lambda>0$, that is, predator is headed to the left of where it appears in the prey's visual field, and dextral otherwise. 

\begin{table}[h]
\caption{Sensory Requirements of Evasion Strategies}
\begin{center}
\begin{tabular}{l | >{\centering}p{0.042\textwidth} >{\centering}p{0.042\textwidth} >{\centering}p{0.042\textwidth} >{\centering}p{0.042\textwidth}| >{\centering\arraybackslash}p{0.065\textwidth} }
            & 
            \multicolumn{4}{c|}{Predator state}
            &
            \\
            \cline{2-5}
            & angular
            &  
            & heading
            &   
            & Complexity \\ 
            & position 
            & heading
            & deviation
            & speed
            & of sensing \\
            & $\phi$ 
            & $\psi$
            & $\lambda$ 
            & $V$
            & \\
            \hline
Distance-optimal     & $\cdot$ & $\bigcirc$  & $\circ$ & $\bigcirc$  &  most \\  
Orthogonal  & $\cdot$ & $\bigcirc$  & $\circ$ & $\cdot$ &  \\ 
Parallel    & $\cdot$ & $\bigcirc$  & $\cdot$ & $\cdot$ &  {$\big\downarrow$}  \\
Antipodal   & $\bigcirc$  & $\cdot$    &   $\cdot$  & $\cdot$ & \\
Contralateral & $\circ$ &  $\cdot$  &  $\cdot$   & $\cdot$  & least \\\hline 
\multicolumn{6}{c}{}\\
\multicolumn{6}{c}{$\bigcirc$ exact value \qquad $\circ$ interval value \qquad $\cdot$ not needed}
\end{tabular}
\label{tab:sensing}
\end{center}
\end{table}
\subsubsection*{Descriptive statistics of kinematic measurements} 
We found no correlation between the prey's escape direction $\theta$ and its distance $d$ from the predator at the onset of evasion (SI, Fig. S3). 
However, we did find a clear correlation between the escape direction $\theta$ and the angular position $\phi$ in instances where the predator appears in the prey's visual field (SI, Fig. S5). The data also showed a correlation between $\theta$ and the predator heading $\psi$, when partitioned based on whether the predator's heading is sinistral ($\lambda>0$) or dextral ($\lambda<0$), relative to its angular position $\phi$ (SI, Fig. S7).  Importantly, although the distributions of $\phi$, $\psi$, and $\theta$  varied with predator speed $V$, the correlations between  $\theta$ and  $\phi$ and between $\theta$ and $\psi$ were qualitatively similar for all $V$ (SI, Figs. S5 and S7), suggesting that for the range of speeds considered, $V$ can be treated as a model parameter, rather than a variable that fundamentally changed the evasion behavior. 

In sum, our statistical analysis (SI, Figs. S2-S7, Table S1) indicates that the escape direction $\theta$ depends on the prey's sensing of the predator's angular position $\phi$, heading $\psi$, and deviation between them $\lambda$, but does not disambiguate which stimuli determine the escape direction
and the behavioral rules that best explain the data.

\begin{figure*}[!t]
\centering\includegraphics[scale = 1]{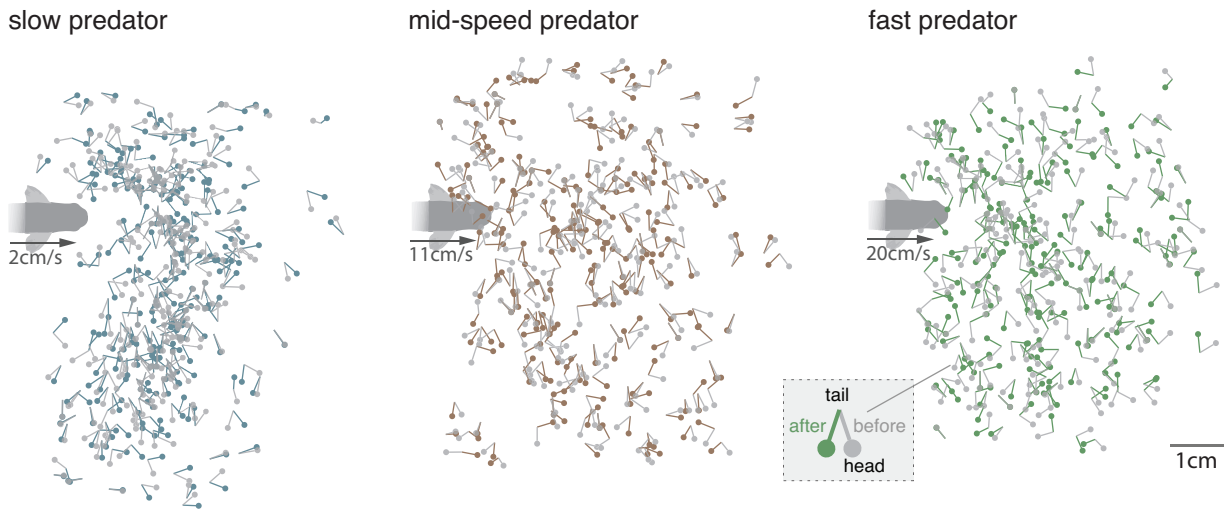}
\caption{Experimental measurements of zebrafish larvae evasion in response to robotic predator.
Zebrafish larvae were randomly placed in a tank with an approaching robotic predator driven at three speeds: $V=2$, $11$ and $20$ cm$\cdot$s$^{-1}$. 
They were mostly straight and motionless until exhibiting a fast C-start evasion response to the predator~\cite{Nair2017,Stewart2014}.
Each experiment involved a single predator-prey encounter. The experiment was repeated to collect three large datasets of size $N_\textrm{slow} = 251$, $N_\textrm{mid} = 233$, $N_\textrm{fast} = 215$ for the slow, mid-speed, and fast moving predator, respectively. 
Evasion instances are superimposed for visualization purposes. 
For each evasion instance, we calculated, in the predator frame of reference, the position and orientation of the prey at the onset of evasion (gray macebells where the head represents the prey's position and spike represents its orientation). The change in prey's orientation $\theta$ induced by the C-start evasion response (see inset) is shown in colored macebells. Color is used only for illustration purposes and not to be confused with the color code used in Figs.~\ref{fig:strategies},~\ref{fig:heatmap},~\ref{fig:modelEval},~\ref{fig:reimaginedEval} to distinguish between evasion strategies.
}
\label{fig:expData}
\end{figure*}

\subsection*{Definition of evasion strategies} 
We next define the six fish evasion strategies: distance-optimal, orthogonal, parallel, antipodal, contralateral, and pure-protean. We index these strategies with an integer $n = 1,\ldots,6$ in the order listed above.
In all strategies, we ignore the prey biomechanics and treat both the predator and prey as point masses equipped with heading directions. 
Therefore, the prey's strategy is demonstrated by the direction of its escape $\theta$.
However, these escape direction vary among the strategies depending on the relative position and heading of predator 
(Fig.~\ref{fig:strategies}B). 
We rate the strategies by their complexity of sensing (Table \ref{tab:sensing}), which is a relative measure that increases with the number of geometric parameters that must be accurately determined to execute the escape in the direction predicted by the strategy.
By this metric, the sensing of exact quantities is more complex than interval quantities.

\subsubsection*{Distance-optimal evasion strategy} A distance-optimal evasion strategy considers that the prey's objective, once it detects the predator, is to maximize its minimum future distance from the predator~\cite{Weihs1984,Soto2015}. Accordingly, the prey should head in the direction $\theta$ relative to its pre-evasion heading (see SI, section 2),
\begin{equation}
{\theta} = f^{(1)}(\psi,\lambda;\chi) = \left[ 
\begin{array}{l l}
\psi - \chi, & \quad \text{sinistral:}~{\lambda} \in [0,\pi),  \\[0.5ex]
\psi + \chi, & \quad \text{dextral:}~{\lambda} \in (-\pi,0),  \end{array}
\right.
\label{eq:optimal}
\end{equation}
 where $\chi= \cos^{-1} (U/V)$ is an angle that depends on the ratio $U/V$ of prey speed $U$ to predator speed $V$. For $U>V$, $\chi = 0$. We treat $\chi$ as a model parameter rather than a variable. This distinction may not be important for the prey, but it is relevant for our subsequent analysis of this strategy.

\subsubsection*{Orthogonal evasion strategy}
We propose a simpler evasion strategy where the prey turns $90^o$  \emph{away} from the heading direction $\psi$ of the predator,
\begin{equation}
\theta =  f^{(2)}(\psi,\lambda)=
\left[ 
\begin{array}{l l}
\psi - {\pi}/{2}, &  \quad  \text{sinistral:} \  {\lambda} \in [0,\pi),  \\[0.5ex]
\psi + {\pi}/{2}, &  \quad \text{dextral:} \ {\lambda} \in (-\pi,0). 
\end{array}
\right.
\label{eq:orthogonal}
\end{equation}
This strategy is equivalent to the distance-optimal strategy in the fast predator limit $U/V\rightarrow 0$, but may determine $\theta$ without the need to sense the predator speed $V$.

\subsubsection*{Parallel evasion strategy} 
For a slow predator $U/V \geq 1$, the optimal strategy is for the prey to reorient itself in the direction of the predator heading, $\theta =   f^{(3)}(\psi) = \psi$, which can be readily deduced by setting 
$\chi = 0$ in~\eqref{eq:optimal}.
The major disadvantage of this strategy is that it could place the predator in the blind spot of the prey's visual field. 

\subsubsection*{Antipodal evasion strategy} Empirical observations~\cite{Eaton1981,Domenici1993} suggest that the prey might follow an antipodal strategy by reorienting its heading $\theta$ in the direction opposite to the angular position $\phi$ where the predator appears in its visual field, without any account for the predator heading $\psi$,
\begin{equation}
\theta =  f^{(4)}(\phi) = 
\left[ 
\begin{array}{l l}
\phi + \pi, \quad & \text{left stimulus:} \ {\phi} \in [0,\pi), \\[0.5ex]
\phi - \pi, \quad & \text{right stimulus:} \ \phi \in (\pi,2\pi).
\end{array}
\right.
\label{eq:antipodal}
\end{equation}

\subsubsection*{Contralateral evasion strategy}  A similar but simpler strategy, called contralateral, was suggested in~\cite{Nair2017} when the prey is approached by the predator from either side. Accordingly, the prey escapes by turning $90^o$ either to the `left' or `right' of its own pre-evasion heading,
\begin{equation}
\theta =  f^{(5)}(\phi) = 
\left[ 
\begin{array}{r l}
-\pi/2, \quad & \text{left stimulus:} \ {\phi} \in [0,\pi), \\[0.5ex]
\pi/2, \quad & \text{right stimulus:} \ \phi \in (\pi,2\pi).
\end{array}
\right.
\label{eq:contralateral}
\end{equation}

\subsubsection*{Pure-protean evasion strategy} The pure-protean strategy suggests that the evasion response $\theta$ is random, independent of the predator state, with a uniform probability of moving in any particular direction. This strategy is best expressed in a probabilistic manner, where the probability density function (PDF) is uniform with equal probability density $p^{(6)}(\theta)=1/(2\pi)$ of obtaining any change in orientation $\theta$.

\subsection*{Testing evasion strategies} 

We developed a method for evaluating evasion strategies in terms of their ability to explain the experimental observations (Fig.~\ref{fig:expData}). 
Although the pure-protean strategy is not supported by our experimental data (see SI, Figs.~S5-S7), 
 the data exhibits some level of randomness, as indicated by the variability in the location of the 
predator at the onset of evasion, but potentially also due to
inherent sensorimotor noise in the prey’s perception of the predator and its
execution of the evasion response.
Our approach accounts for this variation in evaluating, comparing, and ranking 
the hypothetical evasion strategies. 
To emphasize the generality of our approach, we express it in
terms of a generic stimulus $\mathbf{s}$ and response $\mathbf{r}$, without reference to the specific degrees of freedom that these vectors encompass. For the zebrafish larvae, $\mathbf{r}$ is simply $\theta$, but $\mathbf{s}$ varies depending on the strategy; theoretically, it could encompass all or any combination of the variables that define the predator state $d$, $\phi$, $\psi$, $\lambda$ and $V$.

To examine how well the probabilistic strategy models fit the experimental data, we interpreted the latter from a probabilistic perspective. An experimental dataset generates $N$ samples $(\mathbf{s}_i,\mathbf{r}_i)$, $i=1,\ldots N$, from a joint PDF, denoted by $p_o(\mathbf{s},\mathbf{r})$, whose exact form is unknown.
An evasion behavior follows a conditional PDF $p_o(\mathbf{r}|\mathbf{s})=  p_o(\mathbf{s},\mathbf{r})/p_o(\mathbf{s})$, which is related to the joint PDF $p_o(\mathbf{s},\mathbf{r})$ and 
the PDF $p_o(\mathbf{s})$ of stimuli that elicit an escape response via the Law of Total Probability~\cite{Schervish2012}.
Unfortunately, $p_o(\mathbf{s},\mathbf{r})$ and $p_o(\mathbf{s})$ are unknown, and only discrete samples of these PDFs are available from experiments, thus the need for further modeling and analysis.

\begin{figure*}[!t]
\centering\includegraphics[scale = 1]{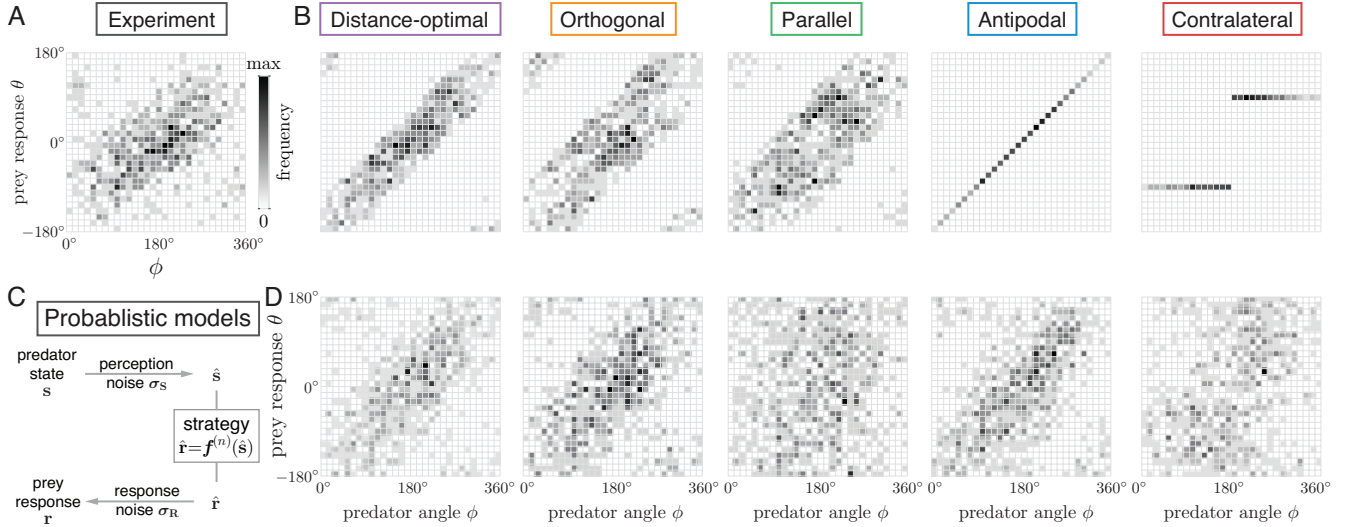}
\caption{Model predictions in response to experimentally observed predator states.  (A) Bivariate histogram of ($\phi,\theta$) from experimental data. Darker color means larger fraction of data points in that area of the $(\phi,\theta)$ space. 
(B) Bivariate histogram based on the evasion models (Eqs~\ref{eq:optimal}--\ref{eq:contralateral}) with no noise;  model predictions $\theta_i$ in response to experimentally observed predator states $\phi_i,\psi_i,\lambda_i$, $i=1\ldots, N$, where $N=699$ is the size of the combined data. (C) Schematic illustration of how noise in sensing and response is built into the evasion models. (D) Bivariate histogram using realizations from the noisy evasion models (Eq. \ref{eq:probmodel}) under optimal noise levels.}
\label{fig:heatmap}
\end{figure*}

\subsubsection*{Probabilistic models under precise vs. noisy sensing and response} We distinguish between the actual predator state $\mathbf{s}$ and the prey's sensing $\hat{\mathbf{s}}$ of the predator state. Similarly, we distinguish 
between the actual escape heading $\mathbf{r}$ and the prey's desired escape heading $\hat{\mathbf{r}}$. 
If the prey's sensing and response are precise, we get $\hat{\mathbf{s}} = \mathbf{s}$ and $\hat{\mathbf{r}}={\mathbf{r}}$.
However, the sensorimotor modalities underlying evasion are often noisy: the prey may perceive a noisy version $\hat{\mathbf{s}}$ of the predator's state $\mathbf{s}$ and its desired response $\hat{\mathbf{r}}$ may be altered by noisy execution or environmental conditions to yield $\mathbf{r}$.

Each evasion strategy $n$, save the pure-protean, defines a desired escape response $\hat{\mathbf{r}}$  given a perceived predatory stimulus $\hat{\mathbf{s}}$ and can be expressed as a conditional PDF using the Dirac-delta function $p\un(\hat{\mathbf{r}}|\hat{\mathbf{s}}) = \delta \left(\hat{\mathbf{r}} - \mathbf{f}^{(n)}(\hat{\mathbf{s}})\right)$. The joint PDF $p\un(\mathbf{s},\mathbf{r})$ formed based on evasion strategy $n$ follows from the Law of Total Probability
\begin{equation}
    p\un(\mathbf{s},\mathbf{r}) = \iint p(\r|\hr) p^{(n)}(\hr|\hs)p(\hs|\mathbf{s})
    p_o(\mathbf{s})\rm{d}\hs\,\rm{d}\hr.
    \label{eq:probmodel}
\end{equation}
Here,  $p(\hs|\mathbf{s})$ and $p(\r|\hr)$ model the noise in the prey's sensing and response. 
In the case of precise sensing and response,~\eqref{eq:probmodel} reduces to
\begin{equation}
    p\un(\mathbf{s},\mathbf{r}) = \delta \left({\mathbf{r}} - \mathbf{f}^{(n)}({\mathbf{s}})\right)
    p_o(\mathbf{s}).
    \label{eq:probmodel_nonoise}
\end{equation}
In the following, we treat each case separately.

\subsubsection*{Evaluating evasion strategies under precise sensing and response}
To obtain samples of the  evasion response predicted by~\eqref{eq:probmodel_nonoise}, 
we use as input the distribution of the empirically-observed stimuli $\mathbf{s}_i$, and we construct a dataset $(\mathbf{s}_i,\mathbf{r}_i^{(n)}= \mathbf{f}^{(n)}(\mathbf{s}_i))$ for each strategy. For each predator speed, we arrive at five datasets representing theoretical predictions of the prey's evasion response according to the distance-optimal, orthogonal, parallel, antipodal, and contralateral strategies. 
Bivariate histograms in the $(\phi,\theta)$-plane for each strategy based on the dataset combining all predator speeds are shown in Fig.~\ref{fig:heatmap}B. 
The histograms represent discrete cross-sections of $p^{(n)}(\mathbf{s},\mathbf{r})$ and can be used to estimate the joint probability of obtaining a predatory stimulus $\phi_i$ and prey response $\theta_i$.
Compared to the histogram obtained from experiments (Fig.~\ref{fig:heatmap}A), the contralateral and antipodal strategies form straight lines because the predicted $\theta_i^{(n)}$ are uniquely determined by the predator angular position $\phi_i$, while the other distributions are spread out due to their dependency on the predator heading $\psi_i$ and $\lambda_i$.

\begin{figure*}[!t]
\centering\includegraphics[scale = 1]{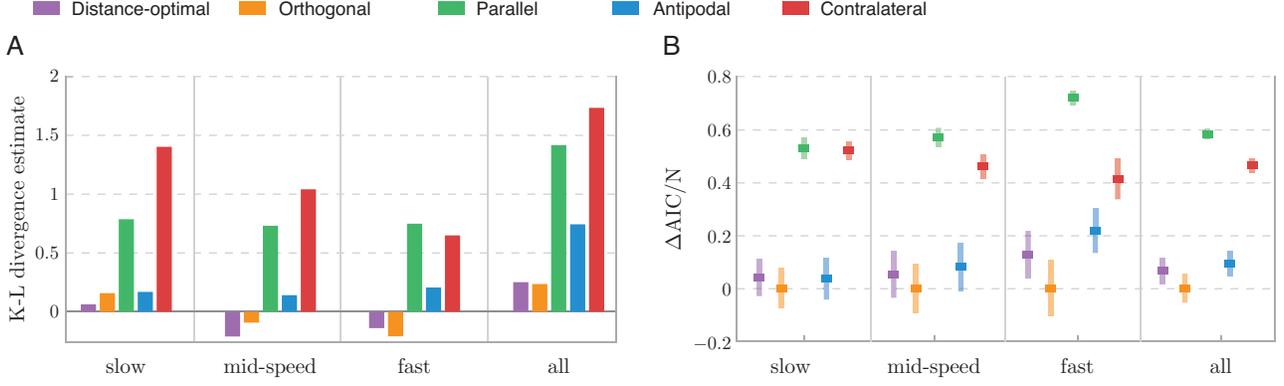}
\caption{Evaluation of precise and noisy evasion strategies. (A) K-L divergence estimate from precise model predictions to experiment data is computed separately for each dataset (slow, mid-speed and fast predator) and for all data combined. The K-L values for the distance-optimal and orthogonal strategies are the lowest, indicating better fit to data. (B) AIC difference ($\Delta$AIC = AIC-AIC$_{\rm{min}}$), normalized by the respective sample size of each dataset.
For each dataset, we used bootstrap method to construct 200 distinct datasets (by sampling with repetition)
of equal size to the original dataset. We optimized each of 200 sets, evaluated the corresponding AIC, and computed the mean and standard deviation of $\Delta$AIC. The orthogonal strategy has the lowest $\Delta$AIC, indicating that it is the most parsimonious strategy and best explains the data.}
\label{fig:modelEval}
\end{figure*}

To measure the difference between model predictions and experimental data, we estimated numerically the Kullback-Leibler (K-L) divergence $\mathcal{D}_{\rm KL}$, which quantifies the entropy of $p^{(n)}(\s,\r)$ relative to $p_o(\s,\r)$,  using the method in~\cite{Perez-Cruz2008}; see SI, S5.
Results of the K-L divergence are shown in Fig.~\ref{fig:modelEval}A for all five strategies applied to the slow, mid-speed, and fast predator, as well as the combined data. The actual K-L divergence is always non-negative; the negative values are due to discrete estimation of the PDF. 
In each of the four datasets, the distance-optimal and orthogonal strategies yield the lowest estimates of the K-L divergence, implying that, of all five evasion strategies, they give the closest predictions of the prey escape response. The distance-optimal strategy performs slightly better for the slow and mid-speed predator while the orthogonal is more advantageous for the fast predator and when considering all data combined. 
The antipodal strategy also gives relatively low K-L divergence estimates.
The parallel and contralateral strategies, whose K-L divergence estimates are significantly higher than the other strategies, have the worst fit to experimental data across all predator speeds.

\subsubsection*{Modeling noise in sensing and response}
We next introduced sensing and response noise according to~\eqref{eq:probmodel}.
To model sensing noise, we considered $\hs$ to be normally-distributed around the actual state of the predator $\mathbf{s}$, with dispersion $\bsigma_\mathbf{S}$, and to model response noise, we considered $\r$ to be normally-distributed around the desired response $\hr$, with dispersion $\bsigma_\mathbf{R}$.  Substituting the noise models $p(\hs|\mathbf{s};\bsigma_\mathbf{S})$ and  $p(\r|\hr;\bsigma_\mathbf{R})$ into~\eqref{eq:probmodel}, and recalling that $p\un(\hr|\hs) = \delta\left(\hr-\mathbf{f}\un(\hs)\right)$, we arrived, for each evasion strategy $n$, at a probabilistic model  that depends on the noise parameters $\bsigma = \{\bsigma_\mathbf{S}, \bsigma_\mathbf{R}\}$ (see SI, section 4). 
Specifically, we used a von Mises distribution (normal distribution on the circle) for $\theta$, $\phi$ and $\lambda$ with noise parameters
$\sigma_\Theta$, $\sigma_\Phi$ and $\sigma_\Lambda$;
we let the noise on $\psi$ follow from $\psi=\phi+\lambda+\pi$ (see SI, section 4).
At zero noise, the von Mises distribution converges to a Dirac-delta function at the mean value; when the noise level is high, it approaches a circular uniform distribution with constant PDF $1/(2\pi)$ in any escape direction. 

\begin{figure*}[!t]
\centering\includegraphics[scale = 1]{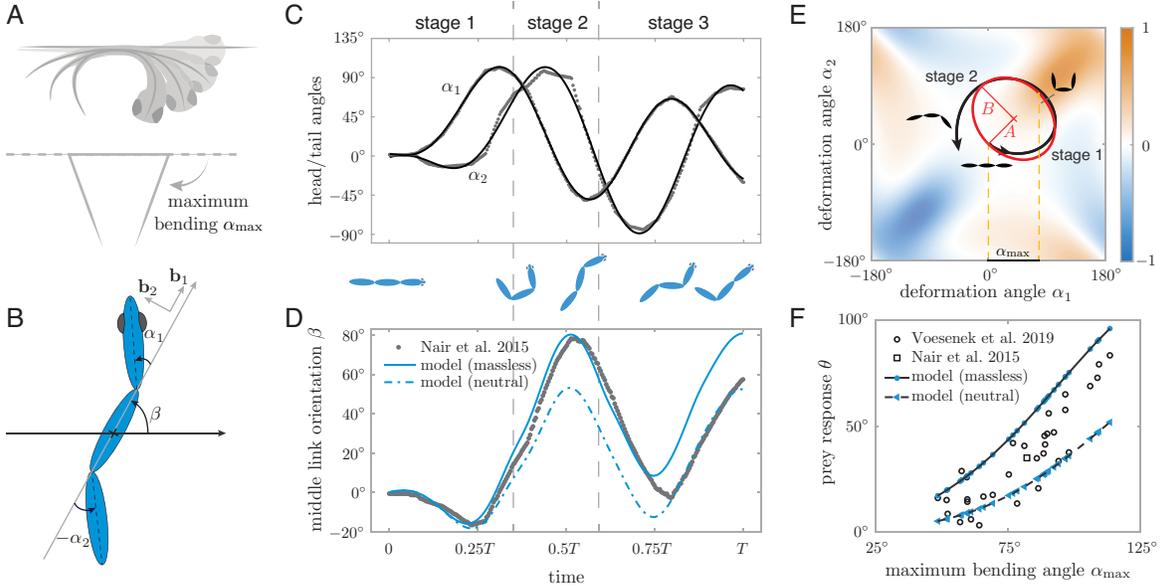}
\caption{\textbf{Biomechanics of fish C-start response.} (A) Larval zebrafish bends its body into a C-shape to initiate a fast start. 
(B) Three fish model, with
 $\alpha_{1}$ and $\alpha_{2}$ representing the fish body shape and $\beta$ the overall body orientation relative to the straight pre-evasion direction. (C) Experimental data of shape changes of larval zebrafish during evasion taken from Ref.\cite{Nair2015} and processed to represent body deformations in terms of head and tail rotations $\alpha_1,\alpha_2$ (gray dots) then fitted by third-order fourier series (black lines). (D)  Experimental data of overall body rotation for the same evasion instance shown in C (gray dots and black line). Predictions based on fish model, taking as input the shape changes in C are shown in blue lines (solid line for massless and dashed line for neutrally-buoyant fish).
(E) The sequence of shape changes in C forms a curve $C$ in the shape space $(\alpha_{1}, \alpha_{2})$ (black line). The experimental curve $C$ is approximated by an ellipse (red) of axes $A,B$ along $\alpha_1=\pm\alpha_2$ directions. Colormap represents curl$_2[A_1,A_2]$ of fish model, which predicts larger turns for curves that encompass solely positive (orange) or negative (blue) values. 
(F) By varying $A$ and calculating $B$ that maximizes the turn, we get a mapping from maximum bending angle $\alpha_{\rm max} = \sqrt{2}A$ to turning angle $\theta$ for massless and neutrally buoyant fish (blue lines) that form an upper and lower bounds on the experimental data set of Ref.~\cite{Voesenek2019}. Both numerical and experimental data show that the C-start mechanics limits larval zebrafish to turning angles $\theta\lesssim100^\circ$.}
\label{fig:3linkFishValidation}
\end{figure*}

\subsubsection*{Limit of high noise levels} 
If the response noise $\sigma_\Theta$ is large, any evasion direction is predicted with equal probability density $1/(2\pi)$, irrespective of the strategy or the sensing noise, that is, all strategies become essentially equivalent to the pure-protean strategy. On the other hand, if the response is precise $\sigma_\Theta=0$, but the noise in sensing the predator's angular position $\sigma_\Phi$ is large, all strategies, except the contralateral, converge to the pure-protean strategy; the contralateral strategy predicts $\theta=\pm\pi/2$ with equal probability. 
If the prey's response and sensing of the predator angular position are both precise $\sigma_\Theta = \sigma_\Phi = 0$, but the prey's sensing of the predator's heading is very noisy ($\sigma_\Lambda$ large), the antipodal and contralateral strategies do not get affected while the parallel strategy becomes protean. Interestingly, in this case, the distance-optimal strategy predicts higher probability of evasion in directions opposite to the predator location spanning a range of $2\chi$ (see SI, section 4, Fig. S9). That is, the distance-optimal strategy becomes a noisy variant of the antipodal strategy. For $\chi = \pi/2$, the orthogonal strategy with large noise on $\lambda$ converges to the antipodal strategy with uniform noise spanning a range of $\pi$ on either $\phi$ or $\theta$. 

\subsubsection*{Optimizing noise levels in sensing and response} For each noisy evasion strategy, we calculated the noise parameters $\bsigma = \{\bsigma_\mathbf{S}, \bsigma_\mathbf{R}\}$ that maximize the total likelihood $\mathcal{L}$ of the model given an experimental dataset, or equivalently  minimize the negative log-likelihood function NLL (see SI, section 6) 
\begin{equation}
    \textrm{NLL} = - \ln \mathcal{L}\left(\bsigma|(\r|\s); n\right) = - \sum_i \ln p^{(n)}(\r_i|\s_i;\bsigma),
\end{equation}
where $p^{(n)}(\r_i|\s_i;\bsigma)$ is the conditional PDF of obtaining a response $\r_i$ given stimulus $\s_i$ for strategy $n$ at noise level $\bsigma$.
The optimal noise parameters $\bsigma^\ast$ are given by 
\begin{equation}
    \bsigma^\ast = \arg \min_{\bsigma} \textrm{NLL}. %
\end{equation}
We solved this optimization problem numerically in the range $\sigma_\Phi, \sigma_\Lambda, \sigma_\Theta \in(0,\pi)$ (SI,~Fig. S6). 
In Fig.~\ref{fig:heatmap}D, we plot realizations generated from the five probabilistic evasion models $p\un(\s,\r;\bsigma^\ast)$ at the optimal noise values corresponding to the dataset of all predator speeds combined.  Compared to the deterministic predictions in Fig.~\ref{fig:heatmap}B, all five distributions appear closer to the experimental data in Fig.~\ref{fig:heatmap}A.

\subsubsection*{Evaluating strategies under optimal noise parameters} To evaluate how well each optimized strategy describes the experimental data, we applied the Akaike information criterion (AIC) defined as \cite{Akaike1974}
\begin{align}
    {\rm AIC} = 2K - 
    2\ln \mathcal{L}(\bsigma^\ast|(\s,\r); n) 
\end{align}
where $K$ is the number of model parameters in each strategy. 
AIC considers both the goodness of fit represented by the likelihood function, and the complexity of the model: if two models have the same likelihood to explain the data, the criterion favors the simpler model.
For example, for the antipodal strategy, we have two noise parameters $\bsigma_\mathbf{S} \equiv \{\sigma_\Phi\}$ and $\bsigma_\mathbf{R} \equiv \{\sigma_\Theta\}$, thus $K=2$; whereas for the orthogonal strategy, we have three noise parameters $\bsigma_\mathbf{S}\equiv \{\sigma_\Phi, \sigma_\Lambda\}$ and $\bsigma_\mathbf{R} \equiv \{\sigma_\Theta\}$, and the orthogonal strategy is deemed more complex than the antipodal strategy.

We used bootstrapping to probe the accuracy of our evaluation of the noisy strategies. Starting from each dataset (e.g., that of the fast predator), we constructed 200 distinct datasets of equal size to the original dataset (e.g., $N_{\rm fast}$) by random sampling with repetition. We solved the  optimization problem 200 times and obtained 200 values of $\bsigma^\ast$ per strategy for each dataset. 
We calculated the likelihood value $\mathcal{L}$ and evaluated the AIC for all 200 optimal noise values, thus obtaining a distribution of AIC values for each strategy and predator speed. The mean and standard deviation of the distributions of AIC values, minus the lowest mean value and normalized by the size of the respective dataset (Fig.~\ref{fig:modelEval}B) show that strategies with lower mean values of the AIC better fit the experimental data.

The results based on the AIC evaluation of the probabilistic strategies in the presence of sensory and response noise are mostly consistent with the results based on the K-L divergence (Fig.~\ref{fig:modelEval}A) for precise sensing and response, but with marked differences. The orthogonal strategy ranks the highest in every dataset; the distance-optimal strategy is slightly behind, in second place, in all but the slow predator dataset where the antipodal strategy ranks second. 
The difference between the orthogonal, distance-optimal, and antipodal strategies is most distinguishable in the case of the fast predator.
The contralateral and parallel strategies come last in all datasets and are least descriptive of experimental data.

\subsubsection*{Further analysis of distance-optimal strategy}

While the predator speed was controlled at $V = 2,11,20$ cm s$^{-1}$,
 the zebrafish larvae  were almost identical  in all experiments, implying that the speed ratio $U/V$ varied drastically between evasion instances: for the fast predator, this ratio is up to 10 times that of the slow predator. If the prey were to sense and use the speed ratio to implement the distance-optimal strategy, we would expect the best performance to appear at different values of $\chi = \cos^{-1}(U/V)$ depending on predator speed. To test this, we evaluated this strategy for the slow, mid, and fast predator as a function of $\chi \in [0, 90^\circ]$ under both precise and noisy sensing and response (SI, section 7, Fig. S14). We found that the K-L divergence  decreased as $\chi$ increased and reached a minimum near $\chi=75^\circ$ independent of predator speed.  Similarly, the NLL dropped as $\chi$ increased until it reached a minimum at, or close to, $\chi=90^\circ$. These results suggest that, even if following the distance-optimal strategy, the prey does not rely on real-time and accurate measurements of the speed ratio $U/V$, but favors the limit of large predator speed ($\chi \to 90^\circ$), where the distance-optimal strategy converges to the orthogonal strategy.

The same conclusion can be reached by examining the values of the optimized noise parameters. 
In the range $20^\circ\lesssim \chi\lesssim 75^\circ$, the optimizer
mostly selects the largest possible value of $\sigma_\Lambda=\pi$ to best fit the data. This high level of optimized noise
indicates that $\lambda$ is not an effective sensory cue in the distance-optimal strategy, and that the prey is unlikely to use this strategy at moderate $\chi$ values   (see SI, section7, Fig. S14).

\subsection*{Evaluating the biomechanical constraints on escape strategy}

To complete our evaluation of fish evasion strategies, we considered the biomechanics of the C-start response.
In~\cite{Nair2015}, the motion of a zebrafish larvae undergoing a C-start maneuver starting from a straight motionless configuration was recorded using high-speed photography, and the time evolution of each segment of the fish body from the onset of evasion at time $t=0$ to after the completion of the C-start response at $t=T=25$ms was measured. 
We developed a mathematical model of the biomechanics of these events and incorporated that model into our analysis.

We reinterpreted the experimental measurements in the context of a three-link fish, head, middle, and tail (Fig.~\ref{fig:3linkFishValidation}B), and we extracted from experimental measurements the fish orientation $\beta(t)$ and rotations $\alpha_{1}(t)$ and $\alpha_{2}(t)$ of the head and tail relative to the middle segment (see SI, section 8-9). 
The time evolution of the zebrafish body during evasion follows the three archetypal stages of the C-start response: in stage 1, the fish curls its body to one side, rapidly unfurls its body in stage 2, and begins its undulatory swimming in stage 3 (Fig.~\ref{fig:3linkFishValidation}C-D).

A larger number of C-start maneuvers were recorded in~\cite{Voesenek2019}, albeit only measuring the maximum degree of body bending $\alpha_{\rm max}$ and the net change in heading $\theta = \beta(T) -\beta(0)$ induced by the C-start maneuver (Fig.~\ref{fig:3linkFishValidation}F).  These results show that the change in heading direction $\theta$ correlates strongly with the degree of body bending~\cite{Voesenek2019}. In all recorded maneuvers, the change in body orientation barely reaches $100^\circ$.

\begin{figure*}[!t]
\centering\includegraphics[scale = 1]{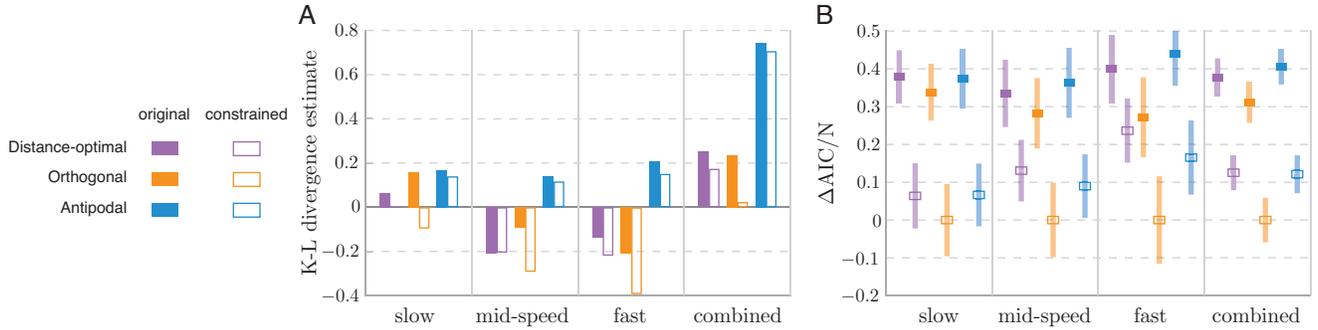}
\caption{\textbf{Evaluation of the constrained strategies that consider the physical constraint on turning.} Results are shown for the three best-performing models. (A) The K-L divergence estimates of the constrained models with precise sensing and response (hollow bars), shown with the results of the original models (solid bars, from Fig.~\ref{fig:modelEval}A). In all four datasets, the constrained models provide discernibly lower K-L divergence estimates, thus better fit to experimental data than the original models, except the distance-optimal strategy for the mid-speed predator. The orthogonal strategy improved the most after imposing the constraint, making it fit the data best in all datasets.
(B) The normalized relative AIC for the constrained models with optimized noise in sensing and response, compared to results using the original models in Fig.~\ref{fig:modelEval}B. The orthogonal strategy still provides best fit to all datasets, marked by the lowest AIC scores, and its advantage over the second best model is more noticeable after imposing the constraint. The constrained antipodal strategy performs comparable to or even better than the constrained distance-optimal strategy.}
\label{fig:reimaginedEval}
\end{figure*}

\subsubsection*{Physics-based modeling of the C-start response} To shed light on the relationship between body deformations and change in heading $\theta$ during evasion, we 
employed a physics-based model of a three-link fish in potential flow~\cite{Kanso2005a,Jiao2021}. Experimental and computational flow analysis had shown that the C-start maneuver is dominated by unsteady, pressure-based exchange of momentum between the fish and surrounding fluid, with negligible contributions from fluid viscosity and shed vorticity~\cite{Gazzola2012,Muller2008}.
The potential flow model captures these unsteady pressure forces via the added mass effect (see SI, S8). 
The fish model is composed of three identical prolate spheroids (of major and minor axes $a$ and $b$)
such that the head and tail are free to rotate relative to the middle link (Fig.~\ref{fig:3linkFishValidation}B); as before, body deformations are described by the angles $\alpha_{1}(t)$, $\alpha_{2}(t)$ representing the relative head and tail rotations as a function of time $t$, and body orientation $\beta(t)$ is the angle between the middle section and an inertial direction taken along the direction of the initially-straight fish.

From consideration of momentum balance on the fish-fluid system, we arrived at an equation governing the rate of change of body orientation~\cite{Kanso2005b,Hatton2010,Jiao2021} (see SI, section 8)
\begin{equation}\label{eq:eom_zero}
 \dot\beta = A_1(\alpha_1,\alpha_2)\dot{\alpha}_1 + A_2(\alpha_1,\alpha_2)\dot{\alpha}_2, 
\end{equation}
where ${A}_1$ and $A_2$ are nonlinear functions of body deformations  $\alpha_1(t),\alpha_2(t)$; 
${A}_1$ and $A_2$ also depend on fish geometry and fluid and body densities ($\rho_{\rm f}$ and $\rho_{\rm b}$). For $\rho_{\rm b}=\rho_{\rm f}$, the fish is neutrally-buoyant. When the fluid forces are dominate, the fish can be considered massless and $\rho_{\rm b}$ is set to zero. Body rotations are proportional to the line integral of \eqref{eq:eom_zero} over a curve $C$ describing body deformations in the shape space $(\alpha_1,\alpha_2)$.  For a closed curve $C$, this line integral can be rewritten, using Stokes theorem, as an area integral over the region of the $(\alpha_1,\alpha_2)$ space enclosed by $C$,
\begin{equation}
\label{eq:rotation}
\theta = \beta(T) - \beta(0) = \int\int \left(\dfrac{\partial A_2}{\partial \alpha_1} - \dfrac{\partial A_1}{\partial \alpha_2}\right) \textrm{d}{\alpha}_1 \textrm{d}{\alpha}_2 . 
\end{equation} 
The scalar field curl$_2([A_1,A_2]) \equiv {\partial A_2}/{\partial \alpha_1} - {\partial A_1}/{\partial \alpha_2}$ is shown in Fig.~\ref{fig:3linkFishValidation}E as a colormap over the entire shape space $(\alpha_1,\alpha_2)$. To maximize the turning angle $\theta$, a straight fish should deform its body following a closed curve $C$ that encompasses either non-positive or non-negative values of  curl$_2([A_1,A_2])$, i.e., either blue or orange regions of the shape space. Closed curves in the orange region lead to turning counter-clockwise. By symmetry, diagonally-opposite curves in the  blue region lead to turning clockwise. Theoretically, the simplest curve for turning is a circle or an ellipse in the shape space  
of major axis A aligned with $\alpha_1 = \alpha_2$,
for which the maximum bending angle is 
$\alpha_{\rm max} = \sqrt{2} A$
(Fig.~\ref{fig:3linkFishValidation}E). 
Corresponding fish shape deformations and body rotations $\beta(t)$ are discussed in SI (S8-9, Figs. S15-S16).

\subsubsection*{Comparing model predictions to C-start induced turning of the fish body} We represented the empirical time evolution of shape deformations $(\alpha_1(t),\alpha_2(t))$  (Fig.~\ref{fig:3linkFishValidation}C) onto the shape space (Fig.~\ref{fig:3linkFishValidation}E). Interestingly, the curve $C$ (black line) traced by the actual fish follows closely the elliptic curve (red line) predicted by the model as best for turning. Moreover, when taking the empirical values of $\alpha_1(t)$ and $\alpha_2(t)$ as input to the physics-based model in \eqref{eq:eom_zero}, the resulting predictions 
of $\beta(t)$ (blue lines in Fig.~\ref{fig:3linkFishValidation}D) follow closely the empirical values of $\beta(t)$ (black line), especially during the first stage of the C-start response, where vorticity is negligible; note that while the neutrally buoyant model (dashed blue line) deviates slightly from empirical observations in stage 2, the massless fish model (solid blue line) performs remarkably well way into stage 3, indicating that indeed unsteady pressure forces dominate the C-start maneuver, as previously predicted~\cite{Gazzola2012}.

We next considered a family of shape changes following the elliptic curve in~Fig.~\ref{fig:3linkFishValidation}E by varying $A$ such that $\alpha_{\rm max}=\sqrt{2} A$ varied from 0 to $120^{\circ}$. 
This upper limit on $\alpha_{\rm max}$ corresponds to a maximum bending angle without causing the head and tail of the model fish to cross each other, and is consistent with the experimental observations of~\cite{Voesenek2019}.
Using~\eqref{eq:rotation}, we computed, for each $\alpha_{\rm max}$, the value of $B$ that optimizes the change in orientation $\theta$, thus creating a map from $\alpha_{\rm max}$ to $\theta$. We compare these model-predictions (blue lines) to experimental data~\cite{Voesenek2019} (black dots) in Fig.~\ref{fig:3linkFishValidation}F. As before, we considered massless and neutrally-buoyant fish. The theoretical predictions behave nearly as upper and lower limits to experimental data. 
As observed previously~\cite{Voesenek2019}, turning in the model fish barely reaches $100^{\circ}$ even when the three-link fish bends its body to the extreme of the head and tail touching.
This indicates that the biomechanics of the C-start maneuver imposes an upper limit on achievable heading directions $\theta$.

\subsubsection*{Constrained evasion strategies}
We next incorporated the physical constraints on $\theta$ imposed by the C-start biomechanics into our evasion strategies. To this end, we mapped the response angle $\theta\un_i$ predicted by evasion strategy $n$ onto the interval $[0, 100^{\circ}]$ using the quadratic mapping 
\begin{equation}
\theta\rightarrow \left[1-\left(1-\frac{\theta_{\rm max}}{\pi}\right)\frac{|\theta|}{\pi}\right]\theta.
\end{equation}
Small turns get less constrained whereas large turns are limited to the maximum angle $\theta_{\rm max} =100^{\circ}$ allowable by the fish biomechanics. We applied this constraint to the three most plausible strategies: distance-optimal, orthogonal, and antipodal. For each constrained strategy, we repeated the analysis presented above under precise and noisy sensing and response. Results of the K-L divergence and AIC analysis for the constrained strategies are shown in Fig.~\ref{fig:reimaginedEval}. 
Compared to the unconstrained strategies,
penalizing large turns makes all three strategies fit better the experimental data across all datasets, with or without added noise, with the exception of the distance-optimal strategy for the mid-speed predator. 
Under precise sensing and response, the relative ranking of the constrained strategies (Fig.~\ref{fig:reimaginedEval}A) is similar to the original ranking  (Fig.~\ref{fig:modelEval}A), with the distinction that the orthogonal strategy at slow and mid-speed predator speed surpasses the distance-optimal strategy and becomes the best ranking model. Under noisy sensing and response, the antipodal strategy ranks higher than the distance-optimal strategy in all but the slow predator dataset (Fig.~\ref{fig:reimaginedEval}B).
Importantly, whether precise or noisy, the orthogonal strategy fits the experimental data better than the other two in all datasets.

\section*{Discussion}

We developed a comprehensive framework for resolving evasion strategy from kinematic measurements.
Our approach considers multiple hypotheses, each defined mathematically (Fig.~\ref{fig:strategies}), that address the role of sensorimotor noise (Fig. \ref{fig:heatmap}) and incorporate the effects of biomechanical constraints (Figs. \ref{fig:3linkFishValidation}--\ref{fig:reimaginedEval}). Importantly, our approach provides a rigorous methodology, rooted in strong-inference principles~\cite{Platt1964}, for revealing the strategy that best fits previous kinematic measurements of zebrafish larvae. This approach eliminates bias towards a particular hypothetical strategy, as done in a previous study that favored the contralateral strategy from the dataset presently analyzed~\cite{Nair2017}.

We found that the responses of zebrafish larvae to evade a predator are best-characterized by the orthogonal strategy (Fig. \ref{fig:modelEval}). This finding challenges the notion that a prey aims either to solely confuse, or maximize its distance from, the predator with its escape \cite{Domenici2011b}. The kinematics of zebrafish do not exhibit the uniform distribution of escape direction $\theta$ characteristic of a pure-protean strategy (SI, Fig. S2E)~\cite{Humphries1970, Driver1988, Moore2017}.
Instead, larvae exhibited correlations between $\theta$ and predator state, including angular position $\phi$ (SI, Fig. S5) and heading $\psi$ (SI, Fig. S7).
The distance-optimal strategy is more predictive of zebrafish kinematics, but is inferior to the orthogonal strategy, based on K-L divergence and the AIC scores (Figs. \ref{fig:modelEval} and \ref{fig:reimaginedEval}).
Therefore, zebrafish larvae do not conform to the classic dichotomy of models for prey strategy.
Although the prevailing patterns favor an orthogonal strategy, variation about the predictions for this hypothesis allows for the possibility of a mixed strategy that could hinder a predator's ability to anticipate the prey's direction. 
These results are relevant to predator-prey encounters, and hence the ecology, of fish species and reflect the advantages and constraints of the prey's neurophysiology and biomechanics.

The distance-optimal strategy requires sensing that may exceed the abilities of larval fish.
This strategy requires detection of the speed of the approaching predator (Table \ref{tab:sensing}), but larval fish possess poor visual acuity, compared to adult fish, due to a relatively small number of retinal cells \cite{Haug2010}.
It has been demonstrated that the escape is triggered by a threshold diameter of a looming visual stimulus, which may be simulated as a circle with an expanding diameter \cite{Dunn2016}.
A looming stimulus alone does not offer the means to differentiate between threats that are small and fast or large and slow. Therefore, the visual system of larval fish may offer a sensory constraint on its ability to perform the distance-optimal strategy.
A more sophisticated visual system could allow for additional cues to gauge the speed or size of a predator, but the processing time necessary to formulate a distance-optimal response may still pose a liability in evasion speed compared with the orthogonal strategy.

The orthogonal strategy merely requires an estimate of the predator's heading and offers tactical benefits relative to many of the alternatives.
This strategy is equivalent to the distance-optimal strategy for a high-speed approach ($U/V\ll 1$) and therefore succeeds in maximizing the prey's distance from a fast predator at reduced sensing requirements (Table~\ref{tab:sensing}).
It is the fastest predators that likely present the greatest threat to the prey. 
The orthogonal strategy offers an additional tactical advantage by evading in a direction that is challenging for a fast-approaching predator to follow because, in order for the predator to execute such large turn at high speed, it needs a large turning radius, which could increase its distance from the prey even further.

The predictions of the orthogonal strategy improved in their fit to measured kinematics when we considered constraints imposed by the biomechanics of the C-start (Fig. \ref{fig:reimaginedEval}).
In particular, our model of a three-link fish in potential flow accurately describes the relationship between the change in fish shape and its turning motion during evasion (Fig. \ref{fig:3linkFishValidation}). 
By mapping maximum bending angle to turning angle, the model predicted an upper limit (around $100^\circ$) on achievable turning motion, consistent with the maximum angle observed in zebrafish exposed to a lateral looming stimulus \cite{Dunn2016, Voesenek2019}.
The improvement in model predictions that included mechanics demonstrates the influence of the constraints imposed by the prey biomechanics and its interaction with the fluid environment on the evasion strategy of zebrafish larvae.

Comparing the K-L divergence and AIC values across slow, intermediate, and fast predators, the prevalence of the orthogonal strategy is clearest in the case of the fast predator (Fig.~\ref{fig:modelEval} and Fig.~\ref{fig:reimaginedEval}).
This feature can be related to the fact that a weak stimulus (slow predator) is more likely to trigger an escape response via the less predictable, long-latency neural pathway~\cite{Burgess2007,Card2012}, as opposed to the fast pathway with minimal latency between perceived danger and motor response~\cite{Walker2005}.
An untangling of these features requires a deeper investigation of how our analytical framework relates to the neurophysiology underlying zebrafish evasion.

Our study combined tools from information theory and probabilistic methods with behavioral evasion models and physics-based models of the C-start biomechanics to develop a comprehensive analytical approach and thereby determine the evasion strategy of zebrafish larvae.
Aside from the details of the biomechanics model, nothing about our approach is specific to the study of fish.
Our analysis could be applied to the myriad of studies that have measured escape responses relative to a predator's approach in a diversity of animals \cite{Domenici2008,Casas2014,Arnott1999,Brown1995,Azizi2002,Woodbury1986}.
This approach may therefore be applied broadly to the study of predator-prey encounters to reveal the strategic basis of this fundamental aspect of animal behavior.

\paragraph{Acknowledgement}
E.K. acknowledges support from the Office of Naval Research (ONR) Grants N00014-22-1-2655, N00014-19-1-2035, N00014-17-1-2062, and N00014-14-1-0421; the National Science Foundation (NSF) Grants RAISE IOS-2034043, CBET-2100209, and INSPIRE MCB-1608744; the National Institutes of Health (NIH) Grant R01 HL 153622-01A1; the Army Research Office (ARO) Grant W911NF-16-1-0074. This research started in summer 2018 at the Summer Graduate School on Mathematical Analysis of Behavior organized by Ann Hermundstad, Vivek Jayaraman, Eva Kanso, and L. Mahadevan. The school was jointly supported by the Mathematical Science Research Institute (MSRI) and the Howard-Hugh Medical Institute (HHMI) Janelia Research Campus, and was held at Janelia. BC, YM, and EK acknowledge support from Janelia and would like to thank Sashank Pisupati and Ann Hermundstad for helpful discussions.

\bibliographystyle{unsrt}
\bibliography{references}

\end{document}